\documentstyle[aps,preprint,tighten,floats,epsf,rotate]{revtex}

\begin{document}
\draft
%%%%%%%%%%%%%%%%%%%%%%%%%%%%%%%%%%%%%%%%%%%%%%%%%%%%%%%%%%%%%%%%%%%%%%%%%

%\preprint{\vbox{\it 
%                        \null\hfill\rm    IP-BBSR/97-43, Nov.'97}\\\\}
%%%%%%%%%%%%%%%%%%%%%%%%%%%%%%%%%%%%%%%%%%%%%%%%%%%%%%%%%%%%%%%%%%%%%%%%%
%
\title{Forming a Single, Large, DCC Domain in a Heavy-Ion Collision}
\author{Sanatan Digal \footnote{e-mail: digal@iopb.res.in}, Rajarshi 
Ray \footnote{rajarshi@iopb.res.in}, Supratim Sengupta 
\footnote{supratim@iopb.res.in}, and Ajit M. Srivastava 
\footnote{ajit@iopb.res.in}}
\address{Institute of Physics, Sachivalaya Marg, Bhubaneswar 751005, 
India}
%
%\date{Oct. 1997}
%
\maketitle
\widetext
\parshape=1 0.75in 5.5in
\begin{abstract}

 We demonstrate the possibility of forming a single, large domain of
disoriented chiral condensate (DCC) in a heavy-ion collision. In our
scenario, rapid initial heating of the parton system provides a driving
force for the chiral field, moving it away from the true vacuum and
forcing it to go to the opposite point on the vacuum manifold. This
converts the entire hot region into a single DCC domain. Subsequent
rolling down of the chiral field to its true vacuum will then lead to
emission of a large number of (approximately) coherent pions. The
requirement of suppression of thermal fluctuations to maintain the
(approximate) coherence of such a large DCC domain, favors three 
dimensional expansion of the plasma over the longitudinal expansion 
even at very early stages of evolution. This also constrains  the
maximum temperature of the system to lie within a window. We roughly 
estimate this window to be about 200 - 400 MeV. These results lead us 
to predict that extremely high energy collisions of {\it very small 
nuclei} (possibly hadrons) are better suited for observing signatures 
of a large DCC. Another possibility is to focus on {\it peripheral} 
collisions of heavy nuclei.

\end{abstract}
\vskip 0.125 in
\parshape=1 -.75in 5.5in
%Key words: Disoriented Chiral Condensate, Pion Condensate, 
%Quark-Gluon Plasma, Thermal Fluctuations}
%\newpage
%\begin{multicols}{2}
\narrowtext
%%%%%%%%%%%%%%%%%%%

\vskip .2in
\centerline {\bf I. INTRODUCTION}
\vskip .1in

 Recently there has been a lot of interest in exploring the very
interesting possibility that extended regions, where the chiral field
is misaligned from the true vacuum, may form in heavy-ion collisions or
in large multiplicity hadronic collisions \cite{anslm,bkt,blz,rw}. Such a
region is termed as a  disoriented chiral condensate (DCC).
If a large DCC domain forms, then it can lead to spectacular signatures
such as coherent emission of pions which can be detected \cite{bkt} as
anomalous fluctuations in the ratio {\it R} of neutral pions to all
pions. A strong motivation for looking for DCC comes from Centauro 
events in cosmic ray experiments \cite{bkt}.

 However, so far it has not been possible to realize this beautiful idea
in the sense of finding experimental situations where large enough DCC
domains can form which can lead to clear experimental signals. Essentially
all theoretical estimates predict rather small DCC sizes, of the order
of couple of fm at best. Not only is the number of coherent pions
very small for small DCC, for a large QGP region undergoing phase
transition, one expects many such DCC domains to form, making it harder 
to experimentally detect the signal from each DCC domain \cite{tpn}.
Though, many clever ideas have been proposed to overcome these problems
\cite{wvlt}, one is still far away from realizing the ideal situation
of having a large single DCC domain with clean experimental signature.

 In this paper, we demonstrate the possibility of forming such a large,
single, DCC domain. In our model, the entire region of the plasma gets 
converted into a single DCC domain, which can have diameter as large as 14
fm for gold nucleus. We work within the framework of linear sigma model,
and study the dynamics of the chiral field using Langevin equations of 
motion to take into account the effects of thermal fluctuations. 
The main idea of our model is that during the rapid heating
of the initial parton system, the chiral field is driven towards the
value zero due to a rapidly changing effective potential. If heating is
fast enough, and the maximum temperature reached is large enough, then
the field picks up enough kinetic energy to overshoot the zero of the 
field and goes to the opposite point on the vacuum manifold. That is, 
starting from the true vacuum with chiral angle being zero, the field 
ends up at chiral angle equal to $\pi$. Most importantly,
this happens for the entire region being heated up. The end result
being that the entire region of the QGP ends up with chiral field being
maximally disoriented, with the chiral angle being (close to) $\pi$.
Eventually the field will roll down from there, emitting coherent pions
in the process. 

  Due to thermal fluctuations, there will be some variation in
the chiral angle over the entire DCC. The requirement that this variation
remains relatively small, compared to the average disorientation of the 
chiral field from the true vacuum, leads to constraints on the maximum 
temperature, as well as on the rate of expansion of the plasma. If the 
maximum temperature $T_0$ of the plasma is too large, or the rate
of expansion too slow (e.g. in the longitudinal expansion model), then 
the system spends significant amount of time at high temperatures so that 
thermal fluctuations become dominant. We find that thermal fluctuations
remain small only when spherical expansion is assumed even at very early
stages of the plasma evolution. For that case we find allowed
values of $T_0$ to lie roughly in the range 200 - 400 MeV, the lower
bound arising due to the fact that for small $T_0$ the chiral field
does not overshoot at all. Such values of $T_0$ are small compared to 
the maximum QGP  temperatures expected in central events in 
the heavy-ion collisions involving very heavy nuclei at LHC and RHIC. 
This suggests that in such experiments, or for that matter, even in 
present heavy-ion experiments, observing a large DCC is more likely
if one explores a range of centrality in selecting events. This is
because with varying centrality the energy density, and hence plasma
temperature, will also be expected to vary. Then the possibility will be 
larger for the plasma temperature to lie in the allowed range for some
events. Requirement of early three dimensional expansion suggests 
that possibility of a large single DCC domain may be larger for very 
energetic collisions of smaller nuclei, or even hadron-hadron collisions. 
In this sense, Tevatron  may indeed be a good place to look for 
DCC \cite{tev}.  This is, in some sense, consistent with what one 
expects from Centauro events where very heavy nuclei are not involved in 
the collision. Precise constraints on the masses of colliding nuclei 
and their energies, in the context of our model, are difficult to 
estimate at this stage. We hope to give somewhat more definite numbers 
in a future work.

 The paper is organized in the following manner. Sec. II discusses
the physical motivations and various assumptions underlying the
basic picture of our model. Sec. III discusses the numerical calculations 
and results. Our results naturally lead to consideration of the effects 
of the topology of field configurations on observables
like pion distribution. This is discussed in Sec. IV. Conclusions
are presented in Sec. V where we summarize our results and discuss
various predictions.

\vskip .3in
\centerline {\bf II. PHYSICAL PICTURE OF THE MODEL}
\vskip .1in

 In this section we will discuss the basic picture underlying our
model. Our numerical technique will be similar to that used in
ref. \onlinecite{lngvn1,lngvn2} and we will describe it in Sec.III.
We will work within the framework of linear sigma model, 
with the finite temperature effective potential at one loop order
given by,

\begin{equation}
V \,=\, \frac{\lambda}{4} (\sum_i \Phi_i^2)^2 - {\lambda \sum_i \Phi_i^2 
\over 2} \left(f_\pi^2 - \frac{m_\pi^2}{\lambda} - \frac{T^2}{2}
\right) - f_\pi m_\pi^2 \Phi_4 \ , 
\end{equation}

\noindent where $\Phi_i$ represents components of the chiral field 
with $\Phi_i = \pi_i$ for $i=1,2,3$ and $\Phi_4 = \sigma$. Here
$T$ is the temperature. We take $m_{\pi}$ = 140 MeV, $f_{\pi}$ = 
94.5 MeV, and $\lambda \simeq 55$ (corresponding to $m_{\sigma} = $ 
1 GeV).  In the chiral limit there is a second order phase transition 
with the critical temperature $T_c = \sqrt{2} f_{\pi}$ ($\simeq 134$ MeV 
for our choice of parameters).   With non-zero quark masses, chiral 
symmetry is explicitly broken, and is only approximately restored at a 
temperature $\simeq$ 115 MeV in the sense that the saddle point at the 
chiral angle $\theta = \pi$ disappears above this temperature 
\cite{kpst1}. In the chiral limit, spontaneous 
breaking of chiral symmetry for temperatures below $T_c$ implies that 
one particular point on the vacuum manifold $S^3$ (characterized by 
$\sum_i \Phi_i^2 = f_{\pi}^2$) will be chosen as 
the vacuum state in a given region of space, with all points on $S^3$ 
being equally likely. One may expect this to essentially hold true even 
in the presence of small pion mass. This will lead to a sort of domain 
structure in the physical space where each domain will have the chiral 
field aligned in a given direction, but the directions in different 
domains vary randomly.

 This essentially summarizes the conventional picture of the formation
of DCC domains. Formation of this type of domain structure in a phase
transition has been extensively discussed in the context of topological 
defects in condensed matter physics and also in particle physics models 
of the early Universe. For an equilibrium, second order phase transition 
one expects \cite{rw} maximum size of domains to be of the order of 
$m_{\pi}^{-1}$. For such small domains, dramatic signals like fluctuation 
in the ratio {\it R} of neutral pions to all pions will not be 
observable. However, it has been suggested that in a non-equilibrium 
transition coherent pion emission may be observable \cite{rw}, see also 
ref. \onlinecite{gavin1} in this context. The amplification of 
long-wavelength pion modes due to parametric resonance, leading to
larger DCC domains, has been discussed in ref.\onlinecite{kaiser}.
The effects of fluctuations
on the growth of DCC domains, in the conventional picture, have been
extensively studied in refs.\onlinecite{cooper,bynsk}. More recently,
similar investigations, for the case of spherical expansion of the
plasma, have been carried out in ref. \onlinecite{lampert}. Especially in
the annealing scenario \cite{annl} it is suggested that reasonably large 
DCC domains of sizes as large as 5 fm may form. The issue of initial 
conditions for these scenarios, appropriate at high temperatures,
has been discussed in refs. \onlinecite{rndrp}, see also ref.
\onlinecite{krzyk} in this context.  

  The most important aspect of all these models is that one starts with
an intermediate chirally symmetric state, at temperature above $T_c$.
As the plasma cools due to expansion, and chiral symmetry
breaks spontaneously, one considers small correlation domains
and studies their growth. This inevitably leads to multi domain
structure of the entire QGP region.

  We propose to choose a different starting point. Just after the
collision of the nuclei or, say, just before it, the chiral field in 
the region is definitely in the true vacuum with the chiral angle equal 
to zero, and is described by the zero temperature effective potential.
As the partons produced during the collision thermalize, eventually
leading to a hot region with a temperature $T_0$ (which will be in the
chirally symmetric phase for large $T_0$), the effective
potential changes to the one appropriate for that temperature. What
happens during this time? It seems very hard to address this question,
though, one can adopt different physical pictures modeling this
intermediate regime. Presumably the most natural approach is to think
of the initial system of partons as being completely out of equilibrium
which then thermalizes due to interactions \cite{wong,kls}. As
thermalization proceeds, the system approaches a state in (quasi)
equilibrium with a well defined temperature, which is the maximum
temperature of the QGP. Thus, in this picture, at any instant during this
intermediate stage, the system does not have a well defined temperature.
However, departure from equilibrium decreases, in the sense of various
distributions gradually approaching the one appropriate for a
system in thermal equilibrium at temperature $T_0$, as partons interact
with each other and thermalize.

  Unfortunately, it is hard to do any calculations within this picture,
as one has to deal with a non-equilibrium system, which is approaching
an equilibrium state. Thus, for intermediate stages, use of finite
temperature effective potential etc. becomes meaningless. Since the
physics of our model is best described in terms of an evolving effective
potential, we do not follow the above approach. Rather, we assume a
simplified picture wherein we think of the initial system of partons
at zero temperature which is getting heated up, finally reaching the
temperature $T_0$ of the QGP. In this picture, at any instant, one has
a well defined temperature which first increases to a maximum value of
$T_0$, and thereafter starts decreasing due to continued plasma expansion.
The second part, that is cooling due to plasma expansion, is common to 
this picture as well as the conventional picture. However, initial heating
assumed in the latter picture may appear confusing as the parton system is
still expanding (in fact, rather fast in the beginning). Initial heating
is actually supposed to represent the thermalization of the parton system.
So, even though plasma is expanding at any time, rate of thermalization
as well as secondary parton production dominates in the beginning,
and leads to a consistent picture that the temperature is rising in the
beginning. Once thermalization is complete, then only effect of expansion
remains, leading to decrease in the temperature. Another way to think
of the initial heating is to realize that we are only interested in
the dynamics of the chiral field. It is clear that the chiral Lagrangian
as given in Eqn.(1) does not account for all the degrees of freedom,
especially at high temperatures. Thus, one can think of the interaction
of these other degrees of freedom with the chiral field as giving rise
to the heating of the system of the chiral field.

  To summarize the picture so far, we think of the initial thermalization
of the parton system as, the system getting heated up to a maximum
temperature $T_0$ which subsequently cools due to {\it continued} plasma
expansion. This translates in terms of the chiral model with a finite
temperature  effective potential with temperature being prescribed as
a function of (proper) time. One can object on using the chiral model at 
very early times \cite{cooper}, and use of finite temperature effective 
potential for very short time scales in the pre-equilibrium phase of the
parton system.  However, we believe that the basic physics
of our model has generic features which do not depend on such
approximations. The spirit of our calculations is that by making these
assumptions, we are able to demonstrate the essential physics of a
novel mechanism for the conversion of the entire QGP region into a
single DCC.

 In order to give a clear picture of the physics underlying our model of
DCC formation, we first consider a simple case with zero pion mass in
the absence of any damping (or thermal fluctuations). Assume
that the system initially is at zero temperature with the chiral field
pointing in some direction on the vacuum manifold denoted by point P in
Fig.1, all such directions having same energy for zero pion mass.
Thick solid curve in Fig.1 denotes the effective potential (in Eqn.(1))
at this stage. Now suppose that the temperature is instantaneously
increased to a value $T_1$ and kept there, where $T_1$ is less than the
critical temperature $T_c$. [For our choice of
parameters, $T_c \simeq 134$ MeV for zero pion mass.]  The dashed curve
in Fig.1 denotes the effective potential at that
temperature. As the temperature has been increased
suddenly, the chiral field will still be having the value shown by
point P in Fig.1. However, now the effective potential being given by
the dashed curve, the chiral field at point P becomes unstable
and starts rolling down towards the minimum of the new effective 
potential. It is clear that if the value of the potential energy at 
point P, with respect to the new effective potential, is larger than 
the height of the central bump then the chiral field will overshoot the 
bump and roll down to the opposite side on the vacuum manifold. From 
Eqn.(1) it is easy to see that this happens when $T_1 > T_c/\sqrt{2}$.
This then disorients the entire region of the plasma where the field is 
undergoing this dynamics. Of course, in this simplified model the chiral 
field  will start oscillating about zero. In realistic case, due to 
cooling of the plasma, as well as due to damping from the expansion of 
the plasma (and from thermal fluctuations, particle production etc.), the 
field will get trapped on one side of the bump, with both sides being 
equally likely if several oscillations happen before the field settles 
down. We will show below that this basic picture works in rather generic 
situations, even with non-zero pion mass, and leads to conversion of the 
entire region (the region which has the same temperature evolution as a 
function of proper time) into a single DCC, apart from the effects of
thermal fluctuations which we will discuss below.

\vskip .3in
\centerline {\bf III.  NUMERICAL CALCULATIONS AND RESULTS} 
\vskip .1in

 We now discuss the realistic case with non-zero pion mass and with
a temperature evolution consisting of initial heating stage with
subsequent cooling.  It is not clear what sort of time dependence is
appropriate for the temperature in the thermalization (heating) stage.
What we will do is carry out calculations for different types of time
dependencies for the temperature during the heating stage, and present
results for all such cases. For the temperature evolution during the
cooling stage, we take the following time dependence \cite{bj}.

\begin{equation}
T(\tau) = T_0 ({\tau_0 \over \tau})^{\eta \over 3} ~ .
\end{equation}

 Here, $\tau_0$ is the proper time at which the maximum QGP temperature
$T_0$ is achieved. We consider the longitudinal expansion model of
Bjorken \cite{bj} for the plasma for which $\eta$ = 1, as well as
the spherical expansion case where $\eta = 3$. What we find is that
spherical expansion is preferred for our model due to requirement of
suppression of thermal fluctuations. Fig.2 shows plots of temperature
as a function of the proper time for the spherical expansion case. 
We have chosen widely different possibilities for the temperature 
evolution during the heating stage, starting from the linear heating, 
shown by the dotted curve, to instantaneous heating with two different 
values of $\tau_0$ shown by the solid, and dashed curves. For the linear 
heating case, we keep the temperature equal to zero up to a value of
$\tau = \tau_{in}$ which sets the scale for the expansion of the volume, 
as discussed below. ($\tau_{in}$ = 0.2 fm and $\tau_0$ = 0.4 fm for 
the dotted curve in Fig.2.)
  
To account for the effects of thermal fluctuations on 
the evolution of the chiral field, we study the dynamics using the 
following Langevin equations of motion\cite{lngvn1,lngvn2}.

\begin{equation}
{\ddot \Phi}_i ~+~ ({\eta \over \tau} ~+~ \eta^{\prime}) {\dot \Phi_i} 
~=~ - {\partial V(\tau) \over \partial \Phi_i} ~+~ \xi_i(\tau) .
\end{equation}  
 
 Here $V(\tau)$ is the effective potential in Eqn (1). 
$\eta^{\prime}$ is the friction coefficient due to coupling to 
the heat bath and $\xi_i(\tau)$ represents Gaussian white noise term with
$<\xi_i(\tau)> = 0$ and $<\xi_i(\tau_1), \xi_j(\tau_2)> = {2T \over 
{\cal V}} \eta^{\prime} \delta_{ij} \delta(\tau_1 - \tau_2)$. Here, 
${\cal V}$ is the volume of the region over which the chiral field is 
being evolved by above equations. As mentioned in 
ref.\onlinecite{lngvn1,lngvn2}, these Langevin equations can be 
interpreted as representing the dynamics of the soft modes (say zero 
mode over volume ${\cal V}$) interacting with thermal fluctuations (heat 
bath) obtained by integrating out the hard modes which are assumed to be 
in thermal equilibrium, see ref. \onlinecite{lngvn,lngvn1} for details. 
In the absence of any thermal fluctuations, the chiral field, everywhere 
in the entire plasma region, $\Omega$, will end up at $\theta = \pi$ after
overshooting. However, in the presence of thermal fluctuations, the
chiral field in different portions in $\Omega$, starting everywhere with 
$\theta = 0$, will overshoot to different points near $\theta = \pi$, 
thereby forming a patch at $\theta = \pi$. We would like to get an 
estimate of the variation of the chiral field in $\Omega$, after 
overshooting, arising from thermal fluctuations. For this, we think of 
the whole plasma region $\Omega$ in terms of smaller volumes ${\cal V}$,
and study evolution of average field in volume ${\cal V}$ with different 
realizations of the noise term in Eqn.(3). Each such realization then 
represents the evolution of the chiral field in one of the volumes 
${\cal V}$ in the region $\Omega$. It is the issue of how the system 
becomes uncorrelated due to thermal fluctuations during a 
non-equilibrium evolution. There is a similarity between this 
and the situation studied in ref. \onlinecite{bynsk} where growth of 
domains was studied for a system undergoing a quench. In our case, the 
situation is reversed, as here one wants to know how correlations break 
up as the system is heated up.

 For the friction coefficient $\eta^{\prime}$, an estimate from ref.
\onlinecite{lngvn,lngvn1} for our case 
(with $\lambda \simeq$ 55) will suggest 
a value of about 40 fm$^{-1}$ (corresponding to the largest value of the
temperature which is 350 MeV in our case , $\eta^\prime$ will be smaller 
for lower temperatures). This value is far too large and the field 
dynamics is extremely damped in this case, and completely dominated by 
thermal fluctuations. For the parameters used in ref. \onlinecite{lngvn1},
the estimated value was about 2 fm$^{-1}$. As mentioned in ref. 
\onlinecite{lngvn1}, such large value of $\eta^\prime$ gave large thermal 
fluctuations, hence small values (by about a factor 1/4) were also used in 
that work which gave smaller thermal fluctuations. In our case when we 
use a value of $\eta^\prime \sim 4$ fm$^{-1}$, which is about 1/10 of 
the estimated value, we get decent results, though still thermal 
fluctuations are somewhat large. As we will argue below, in such case 
formation of large DCC is somewhat less likely, though our scenario of
large DCC formation still works in this case. We find that thermal 
fluctuations are under control when $\eta^\prime$ is taken to be about 1 
fm$^{-1}$ or less. In this case our scenario works very well, and
leads to formation of reasonably coherent, large DCC
domain. Also, the probability of formation of such DCC does not suffer
from extra suppression due to large thermal fluctuations. This
suppression is really small for $\eta^\prime$ = 0.6 fm$^{-1}$ and,
in this paper, we give detailed plots etc. for this value. We will
quote results for the case when $\eta^\prime$ is increased to larger 
values (e.g. 1, 4 and 10 fm$^{-1}$). 

 We now argue  that the estimates of ref. \onlinecite{lngvn,lngvn1} 
should not be rigidly applied for the strongly coupled case of linear 
sigma model and it is reasonable to consider much smaller values of 
$\eta^\prime$ This is because the 
calculations in ref. \onlinecite{lngvn} apply for a weakly coupled theory 
where perturbation calculations can be trusted. This is far from true in 
the present case where the coupling constant is as large as 55. (Of
course, this objection can be raised for any perturbation calculation
for strongly coupled linear sigma model). Even more importantly, as 
clearly stated in ref. \onlinecite{lngvn}, the Markov approximation 
(using which the dissipation term was derived) is valid in the weak 
coupling regime where the relaxation time of the soft modes is much larger
than the collision time. In order that the soft modes have larger 
relaxation time, one can not take $\eta^\prime$ as large as 40 fm$^{-1}$ 
or so.
  
 As estimates of $\eta^\prime$ from ref. \onlinecite{lngvn,lngvn1} for the 
linear sigma model case are not reliable, it is reasonable to explore 
range of values of $\eta^\prime$. This is what we have done. For large 
values of $\eta^\prime$ we get too much thermal fluctuations, while for 
smaller values thermal fluctuations are under control. In fact, one can 
argue that it virtually makes no sense to take $\eta^\prime$ as large as 
40 fm$^{-1}$ in the Langevin equation. 
With such large $\eta^\prime$, second 
order time derivative in the field equations becomes ineffective and the 
situation becomes that of extreme damping. This would not look like any 
relativistic system any more and would resemble more a damped condensed 
matter system. Also, in order that soft modes have at least somewhat 
larger relaxation time than the hard modes (which is the basis of whole 
analysis in ref. \onlinecite{lngvn}) it is important that $\eta^\prime$
should not be much larger than 1 fm$^{-1}$, so that soft modes do not 
thermalize immediately. For example when we express the equation
in terms of dimensionless variables (so, the only parameter that enters
field equations is $\lambda$), we find that for our case, the value of 
$\eta^\prime = 0.6$ fm$^{-1}$ translates to a value about 0.5, whereas a 
value of 40 fm$^{-1}$ will translate to about 32. This, we believe, also 
suggests that such large values of $\eta^\prime$ are not reasonable.

 With these points in mind, we go ahead with using $\eta^\prime = 0.6$ 
fm$^{-1}$.  We will also quote results for larger values of $\eta^\prime$ 
(e.g. 1, 4 , and 10 fm$^{-1}$) for comparison. We will see that 
qualitative aspects of our scenario are not unaffected when we increase 
$\eta^\prime$ to 4 fm$^{-1}$. As thermal fluctuations are larger in this 
case, formation of very large DCC is less likely in this case. For example 
in one estimate we argue that for $\eta^\prime \sim 4$ fm$^{-1}$ the 
probability of a large DCC is about 1/1000 (per event) compared to the 
case when $\eta^\prime \sim$ 1 fm$^{-1}$.

 Volume ${\cal V}$  over which the evolution of the soft modes is 
considered via Eqn.(3), will expand as plasma expands. We take 
${\cal V} = {\cal V}_0$ at $\tau = \tau_0$. For the linear 
heating case, this amounts to taking a smaller value for ${\cal V}$ at 
$\tau = \tau_{in}$, which then expands to become equal to ${\cal V}_0$ 
at $\tau = \tau_0$. The volume expands via the following equation,

\begin{equation}
{{\dot {\cal V}} \over {\cal V}} = {\eta \over \tau} .
\end{equation} 

It may appear reasonable to take ${\cal V}_0$ to be of order of a 
correlation volume. Then, a natural value to take will be ${\cal V}_0 
\simeq 1$ fm$^3$.  (This is for the instantaneous heating case. For the 
linear heating case it will lead to consideration of rather small volume 
at $\tau = \tau_{in}$, especially for the spherical expansion case.)
Again, thermal fluctuations are larger in this case, so we give results 
for a larger value ${\cal V}_0 = 5$ fm$^3$. (The radius of such a region
will be about 1 fm.) Results for even larger value of ${\cal V}_0$ 
further reduce thermally generated pionic components, otherwise results 
are essentially the same. It is important to remember that ${\cal V}$ 
does not really represent a correlation volume. It is a volume 
over which the average value of the chiral field is being evolved. Thus, 
in principle, we can take even larger values of ${\cal V}_0$ (see 
ref.\onlinecite{lngvn1}, for example, where values of ${\cal V}_0$ as 
large as 10, 100, 1000 fm$^3$ have been used). However, we are also 
interested in knowing the inhomogeneity in the chiral field generated due 
to thermal fluctuations during its overshooting to the opposite point of 
the vacuum manifold. In other words, we want to know the structure of the 
patch the chiral field forms near the south pole on $S^3$ in the entire 
region $\Omega$ of the plasma after overshoot. For this, one needs to 
divide the entire region into several domains, and study the evolution of 
the average field in these domains. From this point of view, we should
not choose ${\cal V}$ to be too large compared to the correlation length,
as that will govern the scale over which thermal fluctuations will
generate non-uniformities. For this reason, we let ${\cal V}$ expand
only to a value of about 10 fm$^3$, and freeze ${\cal V}$ to this
value during further evolution. If we allow volume to keep expanding, or 
freeze it at a larger value, the effects of thermal fluctuations are 
further suppressed. 
 
 We have used standard stochastic Runge-Kutta algorithm for solving 
the above Langevin equations which is second order in accuracy for the 
deterministic part of the equation and first order for the noise 
part \cite{stchst}. There still remains the issue of the appropriate
time step size for the noise term. For the case of white noise, as in
our case, it seems clear that time step can not be arbitrarily small.
We use time step size equal to 0.005 fm. We have checked how our results
vary when this time step size is increased and decreased by a factor of 
10.  We find that the results with $\eta^{\prime} = 0$ are unchanged, 
so the deterministic part of the equation is not much sensitive to the 
time step size, as one would expect (as long as the time step size is not 
too large to cause error in the evolution by the deterministic part). 
However, with $\eta^{\prime} \ne 0$, the results depend more sensitively 
on the choice of time step size because of the noise term. Though, 
happily, we find that our main results
do not change much, even when we vary time step size by such large 
amount. We find that the thermally generated pionic components, before
the chiral field is de-stabilized by the appearance of the saddle point,
do not change much when we vary the time step size. Subsequent evolution
of the chiral field varies significantly with such large variation of 
time step (we re-emphasize, for $\eta^{\prime} = 0$ entire evolution
had insignificant variation). However, this is easy to understand, as
even a small change in the pionic components, near the saddle point,
can lead to large variation during the subsequent rolling down of the
chiral field on the vacuum manifold. In general, it seems that
the choice of time step size can affect the results due to the assumption
of white noise. In such situations, one will need to
choose an appropriate value of the time step, presumably motivated
by physics of the problem. For example, one may need to keep the
time step size very small in order to get accurate evolution by the
deterministic part of the equation. However, the noise term may be
updated over a larger time step which could be motivated from the
intrinsic scales in the problem.
 
 For the initial conditions, we choose the chiral field to be in the 
true vacuum at zero temperature, at $\tau = 0$, with the chiral angle 
$\theta = 0$, 
that is, $\Phi_{1,2,3} = 0$ and $\Phi_4 = f_\pi$. ${\dot \Phi_i}$ is 
taken to be zero initially. We will give plots for the case of spherical
expansion of the plasma as thermal fluctuations are too large for
the case of linear expansion. For some cases we will also quote results 
for the linear expansion case. The evolution of $\sigma$ field is shown 
in Fig.3 corresponding to temperature evolutions shown in Fig.2. In 
two of the cases we see, in Fig.3, that $\sigma$  field overshoots the 
central barrier and ends up becoming {\it maximally} disoriented with 
the chiral angle $\theta$ being (close to) $\pi$, (eventually rolling back 
to $\theta = 0$), e.g. the solid curve corresponding to the
instantaneous heating case, with $\tau_0 = 0.4$ fm.
For $\tau_0 = 0.2$ fm, (again, for instantaneous heating, with the same 
value of $T_0$ = 350 MeV), the field does not overshoot as shown by the
dashed curve in Fig.3 and the field settles to $\theta = 0$ without ever 
crossing the central barrier. This happens due to a faster cooling 
rate in this case, so that the force on the chiral field due to changing
effective potential is operative only for a small time. Sometimes,
the field settles down back at $\theta = 0$ after crossing the central 
barrier more than once, as shown by the case corresponding to the linear
heating case with $\tau_{in} = 0.2$ fm and  $\tau_0$ = 0.4 fm (dotted
curve). DCC formation can happen even in such cases, 
due to additional sources of dissipation in the field dynamics arising 
due to coupling to other fields, as well as due to particle production 
etc. We mention here that in our model $T_0$ can, in principle, be 
below the critical temperature itself (see, ref.\onlinecite{dcctg} in 
this context). For example, for the instantaneous heating case with
$\tau_0 = 3.0$ fm (which certainly is a rather large value) we find 
that the field overshoots to the opposite point on the vacuum manifold 
with $T_0$ as small as 107 MeV. This we get for the case of longitudinal
expansion, and in the absence of thermal fluctuations, i.e. 
$\eta^{\prime} = 0$. With non-zero $\eta^{\prime}$,
thermal fluctuations are too large for such a large value of $\tau_0$.
(For the spherical expansion case, we find the overshooting happens
with $T_0$ = 114 MeV for $\tau_0$ = 5 fm.) 

We emphasize here that our interest in a large DCC is from the point of 
view of its experimental signature in terms of measuring the ratio $R$. 
Thus it suffices that at every point in the physical space corresponding 
to a DCC, the chiral field rolls down along the same direction on $S^3$, 
even if the field has some spatial/temporal variation. In fact, the 
experimental signature of DCC in terms of the distribution of the ratio 
$R$ can not differentiate a DCC which has essentially uniform value of the
chiral field, from a DCC where the chiral field has some spatial/temporal 
variation (still forming a condensate) but rolls down
approximately along the same path on the vacuum manifold $S^3$. One
could distinguish the two cases if one could measure the coherence
of pions coming out of the decay of the DCC. 

  For each set of parameters, we generate an ensemble of trajectories
and find out the fraction of trajectories which overshoot the central
bump at some temperature T $<$ 115 MeV (which is the point when the saddle
point first appears). These are the trajectories which show how the
chiral field gets disoriented from the true vacuum
in the respective elementary domains in the region
$\Omega$. Due to thermal fluctuations (with $\eta^{\prime} \ne 0$), pionic 
components become non-zero during the overshoot and the chiral field 
typically ends up at some point different from $\theta = \pi$. For
the whole region to look like a single DCC, we require that the 
variation of the chiral field in the DCC arising due to thermal 
fluctuations remains small. We do not want to constrain pionic 
components which grow due to rolling down the valley after overshooting
as that corresponds to the evolution of DCC. Thus we require that 
pionic components at the stage when the temperature is less than about 
115 MeV, remain less than about 13 MeV (which is half of the shortest 
radius of the valley at that stage, i.e. the value of $\sigma$ at the 
saddle point with chiral angle equal to $\pi$), making sure that the
average value of $\sigma$ for the trajectories overshoots the central bump. 

This cut off for the magnitude of pionic 
components is somewhat ad hoc. However, we would like to mention that
the value of this cut off is only relevant in determining the allowed
range of $T_0$. This range will be somewhat larger for a larger value
of the cut off, but does not depend on it too sensitively. 
Fig. 4 shows parametric plot of the magnitude of ${\vec \pi}$ vs. 
$\sigma$, showing the evolution of the chiral field in the presence 
of thermal fluctuations.  Here we have plotted different trajectories, 
difference arising due to the random nature of the noise term in 
Eqn.(3), corresponding to the situation with temperature evolution 
shown by the solid curve in Fig.2 (for which $\sigma$ evolution is 
shown by the solid curve in Fig.3). As discussed earlier, these 
different trajectories should be thought of as representing the 
evolution of the (mean) chiral field in different portions, each of 
volume ${\cal V}$, in the entire plasma region $\Omega$. 

 This case represents the situation of a reasonably coherent DCC. Here, 
pionic components are dominated by thermal fluctuations upto a point 
where $\sigma \simeq -26 MeV$. This is also the stage around which
the temperature takes the value of about 115 MeV, when the saddle 
point appears, making the chiral field unstable. At this stage one can 
say that DCC has formed. For later times, i.e. for $\sigma$ less than 
this value, the roll down of the chiral field starts dominating which 
is the situation of the evolution of the DCC. It is important to
realize that much of the apparently erratic behavior of a given
trajectory, in Fig.4, is not due to thermal fluctuations. It arises 
due to rolling down on the vacuum manifold, and due to oscillations
about the valley during the course of roll down. Note that we plot
$\sigma$ vs. $|{\vec \pi}|$. In this case we see that the average 
value of the pionic components due to thermal fluctuations at the 
formation stage of DCC, before the roll down of the chiral field starts 
dominating, remains less than about 13 MeV. This value is much smaller 
than the radius of the valley in the effective potential which has a 
value of about 26 MeV in $\theta = \pi$ direction, and a value of about 
51 MeV in $\theta = 0$ direction. More importantly, the area spanned 
by these thermally generated pionic components at this stage leads to
a patch with area less than 3 \% of the total surface area of the vacuum 
manifold at that stage, i.e. when the temperature is about 115 MeV.

  Fig.5 shows the scatter plot of the values of the pionic components at 
$T = 115$ MeV, when the saddle point at $\theta = \pi$ first appears. 
This scatter plot shows the projection of the values of the pionic
components in the $\pi_1 - \pi_2$ plane for 1000 trajectories. As we 
mentioned, the area of this patch in $\pi_1, \pi_2, \pi_3$ space is 
about 3 \% of the surface area of the vacuum manifold at that stage, 
which we approximate by taking the average of the areas of the 3-spheres 
with the smallest and the largest radii of the valley in the effective 
potential. This DCC then represents a large region in the physical space 
where the chiral field is (approximately)
maximally disoriented, with the spatial variation of the chiral field
in the entire region remaining relatively small. We mention here that 
evolution of individual regions with few fm size in $\Omega$ via 
Eqn.(3), neglecting spatial variations, should be justified only
up to $\tau \sim$ few fm or so. For larger times, coupling between
different domains should become important. For a proper, large time 
evolution, one needs to carry out a full simulation of the field 
evolution in $\Omega$. 
 
Thus we see that in our model, $\theta$ distribution in the DCC typically 
represents a small patch ${\cal S}$ on the vacuum manifold $S^3$ at 
$\theta = \pi$, i.e. near the south pole on $S^3$. If this patch does 
not cover/enclose $\theta = \pi$ then the chiral field in the entire DCC 
should eventually roll down roughly along one direction (strip) on 
$S^3$, leading to emission of (approximately) coherent pions, just as 
will be expected for any large DCC. Let us estimate the probability of 
forming such a patch near $\theta = \pi$ whose 
evolution everywhere in $\Omega$ leads to pions with roughly the same 
value of ratio $R$ of neutral pions to all pions. Let us parameterize
chiral field on $S^3$ by $(\sigma, \pi_3, \pi_2, \pi_1) \equiv
(cos\theta, sin\theta cos\phi, sin\theta sin\phi sin\chi, sin\theta
sin\phi cos\chi)$. By taking the number of pions to be proportional
to the square of the corresponding field component, it is easy to 
show that $R = cos^2\phi$, see ref.\onlinecite{rw}. For a 
rough estimate, divide range of $\phi$ (from 0 to $\pi$) in four equal 
parts. It is clear that if the chiral field in the patch ${\cal S}$ 
lies in any two parts, symmetric about $\phi = \pi/2$, and its evolution 
remains confined to these same parts, then it should lead to roughly same 
average value of $R$ for emitted pions. The probability for this type of 
patch ${\cal S}$ to form is  roughly $({1 \over 2})^{(N-1)}$, where $N$ 
is the number of individual domains in $\Omega$. Probability for other 
parameterizations of $S^3$ will be same. Since we freeze the volume
${\cal V}$ after initial expansion at a value of 10 fm$^3$, the 
probability of forming a DCC with 100 fm$^3$ volume will be about 0.002. 
For a 200 fm$^3$ DCC this probability is about $10^{-6}$. For larger 
${\cal V}$ probability for a large DCC is significantly higher. For
example, if ${\cal V}$ is frozen after expansion to a value of 20 fm$^3$,
then the probability for a 200 fm$^3$ DCC will be about 0.002. It is 
possible to argue that similar evolution is expected to happen also when 
the patch covers $\theta = \pi$ but is very asymmetric about it. Clearly, 
this can lead to a much larger DCC. A careful study 
of DCC evolution in our model, with appropriate boundary conditions
at the surface of $\Omega$ where chiral field is fixed at $\theta = 0$,
requires numerical simulation of the chiral model in 3+1 
dimensions, and we plan to present it in a future study. 
It is clear that generic situation will be when the patch is 
reasonably symmetric about $\theta = \pi$ after the overshooting. This
leads to some interesting possibilities and we will discuss it in detail 
in the next section. 

 In the above estimate of the probability of DCC formation we did not
account for the fact that due to thermal fluctuations the trajectories
in some of the elementary domains in $\Omega$ may not overshoot the
central bump (after the time the temperature drops below 115 MeV). We 
have checked that with with $\eta^\prime = 0.6$ fm$^{-1}$, essentially all 
($ 99 \%$) trajectories overshoot the central barrier after temperature 
drops below 115 MeV, so there is no extra suppression in the probability 
of DCC formation. With $\eta^\prime$ = 1 fm$^{-1}$ also, almost all
(97 \%) trajectories overshoot. When we repeated the simulation for a 
larger value of $\eta^\prime$ (= 4 fm$^{-1}$), we find that thermal 
fluctuations are larger and a smaller fraction ($\simeq 45 \%$) of 
trajectories overshoot the central bump. (In all these cases we keep 
$\tau_{in}$ = 0.4 fm and $T_{max}$
= 350 MeV, with instantaneous heating). This will suppress the probability 
of formation of large DCC. For example, in the above estimate of the 
probability, we should multiply by a factor of about 1/2 for
each elementary domain (in the region $\Omega$) to account for the
probability that the trajectory in that domain may not overshoot
the central barrier. This means that for a DCC of volume 100 fm$^3$,
with each elementary domain having 10 fm$^3$ volume, the probability
will be suppressed by an additional factor of 10$^{-3}$. 
Also, due to larger thermal fluctuations in this case, at the time 
trajectories overshoot the central barrier, the average pionic components 
form a patch of area which is about 27 $\%$ of the area of the vacuum 
manifold. This represents a DCC which is not very coherent, though its
subsequent evolution may still be similar to a DCC.

 When we increase $\eta^\prime$ to a value as large as 10 fm$^{-1}$ (which
is about 1/4 of the estimate based on ref.\onlinecite{lngvn,lngvn1}), we 
find that the patch formed by pionic components (at the time of
overshooting the central bump) is too large, covering almost half of the
vacuum manifold and destroying the coherence of DCC. Also, now a very 
small fraction, about 8 \%, of the 
trajectories overshoot the central barrier after temperature drops to 
115 MeV which will lead to strong suppression in the probability of 
formation of large DCC.

 We have also checked how these results change if we take different
set of parameters. For example, for the set of parameters used in
ref. \onlinecite{lngvn1} (with $\lambda$ = 20), we find that $\tau_{in}$
= 0.9 fm and T$_{max}$ = 280 MeV is preferred. With these, about 40 \%
trajectories overshoot the central barrier for
$\eta^\prime$ = 0.4 fm$^{-1}$, with pionic components at overshoot
forming a patch of area which is about 15 \% of the area of the vacuum
manifold. For larger $\eta^\prime$, the fraction
of trajectories overshooting becomes smaller (about 20 \% for 
$\eta^\prime = 1$ fm$^{-1}$). 

  Having discussed the effects of larger values of $\eta^\prime$, let 
us now discuss constraints on the value of $T_0$ in our model. 
For small $T_0$, the chiral field does not pick up enough energy to 
overcome the central barrier. We find the lowest allowed value
$T_{min}$ of $T_0$ to correspond to the case of linear heating
with $\tau_0 =$ 0.6 fm, and $\tau_{in}$ = 0.3 fm. For this case
$T_{min}$ is found to be about 225 MeV. With $\tau_0$ = 0.5 fm
we find $T_{min}$ to be about 240 MeV for the linear heating
case, (for the instantaneous heating case with $\tau_0 = 0.5$ fm we
find $T_{min}$ = 260 MeV). $T_{min}$ increases with decreasing 
value of $\tau_0$. For values of $\tau_0$ larger than 0.6 fm, 
thermal fluctuations become dominant and there is no allowed range
for $T_0$. 

  Upper bound $T_{max}$ on $T_0$ arises from the requirement that thermal
fluctuations do not dominate, so the entire region still looks like an 
approximate DCC. For this we impose requirement that thermally generated 
pionic components remain relatively small. As discussed earlier, we do 
not want to constrain pionic components which grow due to rolling down 
the valley after overshooting. Thus we require that pionic components at 
the stage when the temperature is 115 MeV, remain less than about half of
the shortest radius of the valley at that stage which is about 26 MeV.
Largest value of $T_{max}$ is found to be about 390 MeV,
for the case of instantaneous heating with $\tau_0$ = 0.4 fm. With
$\tau_0$ = 0.5 fm, $T_{max}$ is obtained to be about 310 MeV (again,
for the instantaneous heating case). $T_{max}$ decreases with increasing
value of $\tau_0$. Values of $\tau_0$ smaller than 0.4 fm appear
not very realistic, hence we do not consider such small values. Thus,
finally we get the allowed range for $T_0$ to be roughly 225 - 390 MeV.
The plots in Fig.4 and Fig.5 have been given for the case of 
instantaneous heating case with $\tau_0 $ = 0.4 fm. The allowed range of 
$T_0$ for that particular case is about 290 - 390 MeV.

These values of $T_{max}$ are significantly 
lower than the value of the QGP temperature expected in central events at 
heavy-ion collisions at LHC and RHIC involving heavy nuclei. This 
suggests that in order to look for spectacular signals of DCC in these
experiments one should analyze events with a range of values of impact 
parameter, i.e. lower multiplicity events, as the QGP temperature for
large impact parameter collision will be expected to be smaller. Smaller
nuclei should be more appropriate from this point of view. In fact, it 
seems that AGS and SPS may be better suited for large DCC, especially 
if one analyzes events with a range of centrality. In this way some
events may have $T_0$ falling in the right range for DCC formation. 
However, the requirement of early spherical expansion will prefer
$p-p$ ($p-{\bar p}$) collisions for our scenario to be effective.
In this sense, Tevatron may indeed be a good place to look for DCC 
\cite{tev}. This seems entirely consistent with the observation of 
Centauro events where heavy nuclei are not expected to be involved. 
We may mention here that even for the conventional models of DCC 
formation, where DCC domain sizes are expected to be small, it seems 
better to probe peripheral events. This is because for peripheral events, 
resulting QGP region will be expected to have small volume. This will 
allow very few DCC domains to form in a single event, hopefully making it 
easier to analyze signals of DCC pions. Importance of peripheral
events has also been emphasized in ref. \onlinecite{prph} based on
somewhat different considerations.

\vskip .3in
\centerline {\bf IV. TOPOLOGY OF FIELD CONFIGURATION AND PION 
DISTRIBUTION} 
\vskip .1in

 Let us now discuss the case when the patch formed near the south pole
on $S^3$, after overshooting of the chiral field, is reasonably symmetric.
As we mentioned earlier, this type of situation will be expected
to be most generic. It is important to note that even when thermal 
fluctuations are much more dominant than considered above, this type
of field configuration will still form. This is essentially because
larger thermal fluctuations will only increase the size of the patch
at the south pole (for example in the case of $\eta^\prime$ = 10 
fm$^{-1}$, as discussed above). This is very important, 
especially in view of the uncertainties in the estimates of $\eta^\prime$ 
(as discussed in the previous section). As we showed earlier, for large 
values of $\eta^\prime$, formation of a reasonably coherent, large DCC 
becomes less likely.  However, even in that case, our considerations of 
this section will apply. 

 When the patch near the south pole is reasonably symmetric (after 
overshooting) then we expect in this case that the field will roll back 
towards the north pole, which is the true vacuum, in different 
directions. This will lead to  covering of entire $S^3$ once, and hence, 
will lead to the formation of a single large Skyrmion like configuration 
(due to fixed value $\theta = 0$ 
outside $\Omega$, see ref. \onlinecite{kpst2}). As such it does not 
look  anywhere like the standard picture of evolution of a large 
DCC. However, we now argue for the possibility that the evolution of 
such a configuration leads to spatial distribution of pions  which
can be grouped together in terms of definite values of $R$, thereby 
directly probing disoriented condensates of the chiral field.
With the parameterization of chiral field used above, $R$ depends 
only on angle $\phi$ on the vacuum 
manifold ($R = cos^2(\phi)$). 

Now, given the situation that we 
have a (very) large Skyrmion like configuration in the region $\Omega$ 
(possibly of size 15 fm or so), the configuration should evolve to a 
spherically symmetric configuration of standard Skyrmion \cite{kpst2} 
during the process of its collapse, for energetic reasons. This
reasoning is supported by numerical simulations of texture formation
and collapse in the early Universe \cite{txt}. Textures are just
like Skyrmions, apart from the effect of pion mass term. One can argue
that the pion mass term should not affect these considerations of
texture (Skyrmion) collapse. In ref. \onlinecite{txt}, it was found
that a texture of generic initial configuration becomes spherically
symmetric, and approaches the analytical solution, during collapse. 

 Let us assume the region $\Omega$ or, more precisely, a smaller region
containing the collapsing Skyrmion like configuration, to be spherical 
(which should be fairly 
justified for late stages of evolution) with $\alpha$ and $\beta$ 
being the polar and azimuthal angles on the surface of $\Omega$ 
(which is a two sphere $S^2$). Let us further assume that the Skyrmion 
configuration corresponds to angles $\theta,\phi,\chi$ on the chiral 
vacuum manifold $S^3$ varying spatially so that $\phi = \alpha, 
~\chi = \beta$ and $\theta$ varying radially outward from $\theta = 
\pi$ at the center of $\Omega$, to $\theta = 0$ outside $\Omega$, 
which is the true vacuum. With such a Skyrmion configuration, we 
see that the chiral field in a conical shell in $\Omega$, corresponding 
to constant value of angle $\alpha$, has fixed value of $\phi$ and 
hence corresponds to the same value of the ratio $R$. 

 If we assume that DCC pions 
will come out predominantly radially from $\Omega$ due to
rapid expansion of the region, which does not seem unreasonable as
these pions should also have very small momentum in the rest frame
of the plasma, then we simply need to collect pions at the detector
with fixed polar angle $\alpha$. Since one does not know, a priori,
about the orientation of the Skyrmion configuration in $\Omega$, one 
should analyze pions in conical shells with fixed $\alpha$ and with 
varying choices of the north pole of the two sphere in a $4\pi$
detector around the collision point. It is easy to see that this will
also take care of other parameterizations of the chiral field which are
different from the one used in the above argument. For one particular 
choice of the north pole one will start getting pions with $R$ varying as
$cos^2\alpha$, for each choice of circle on $S^2$, say, in a small
interval $\bigtriangleup \alpha$ at fixed $\alpha$. This analysis can be
suitably adopted for fixed target experiments. We would like to mention
that this technique of analyzing pion distribution seems interesting by 
itself. If there is any axially symmetric phenomenon happening in 
heavy-ion collisions which leaves imprints on the distribution of 
low $p_T$ pions, then this analysis may be able to detect it.

\vskip .3in
\centerline {\bf V. CONCLUSIONS} 
\vskip .1in

 We summarize our main results. We have considered the dynamics
of chiral field during very early stages of the evolution of the
plasma. We model the early, pre-equilibrium, evolution of the system
in terms of finite temperature effective potential, with the
temperature first rising to a maximum value $T_0$ (the heating stage), 
and then decreasing due to continued plasma expansion. Due to change 
in the effective potential during the heating stage, the chiral
field is pushed over to the opposite point on the vacuum manifold,
thereby maximally disorienting the chiral field in the entire
plasma region. This should convert the entire plasma region into a 
single DCC domain. Thermal effects tend to create fluctuations in
this DCC region. The requirement that such fluctuations do not become
dominant, so that the whole region still approximates a single
DCC domain, puts constraints on the rate of plasma expansion,
as well as on the maximum allowed value of $T_0$. We find that
spherical expansion is preferred over the longitudinal expansion.
During very early stages of the expansion, for times much less than
the transverse dimension, it does not seem reasonable to invoke 
spherical expansion. Thus for heavy nuclei, formation of such large
DCC domain looks less likely. From this point of view, hadron-hadron
collisions should be a good place to look for the formation of
large DCC since there spherical expansion will be expected to dominate
from rather early stages. 

  Even for the spherical expansion case, if $T_0$ is too large then
the system spends too much time at high temperatures, again leading
to large thermal fluctuations. We find that thermal fluctuations
remain relatively small only when $T_0$ remains less than about 390 
MeV. There is also a lower bound on $T_0$ arising due to the fact that
for small $T_0$ the chiral field does not overshoot at all. We
estimate this lower bound to be about 225 MeV. There are uncertainties
in the determination of these bounds, for example due to the value of 
cutoff of pionic components at overshoot etc. Conservatively, one can
say that there is a window for $T_0$ from about 200 MeV to about 400 MeV, 
which allows for the formation of such a large DCC. We emphasize again 
that these values are rough estimates. Evolution of the field in one 
region of the plasma will be affected by the evolution of the field in
the neighboring regions. This requires a full 3-dimensional
simulation, with proper boundary conditions. 

  We have also proposed an interesting possibility for the field
evolution in the entire plasma region. Here the field, after
overshooting, evolves to form a topologically non-trivial structure 
which looks like a large Skyrmion. We then argue that the collapse of 
this structure should lead to a very definite type of spatial 
distribution of pions. By proper grouping of pions, this should
lead to definite values of the ratio $R$ arising in a predictable
manner in a single event. We suggest that this technique of analyzing
particle distributions could be fruitful, not just in the context
of our model, but in uncovering any axially symmetric phenomenon
in these experiments which leaves imprints on particle multiplicities.

 We mention that the ideas discussed here have strong implications for 
condensed matter systems. One example is the formation of defects via a 
new mechanism, discussed in ref. \onlinecite{dgl}, where field 
oscillations and subsequent flipping (which is just like the chiral field 
overshooting to the opposite point on the vacuum manifold) leads to defect 
formation. Thus, if a small portion of superfluid helium, or 
superconductor, is heated rapidly to a temperature less than the critical
temperature, then  it may lead to topological defect formation, without 
ever going through a phase transition. We will present these results in 
a separate work.

\acknowledgments
 We are thankful to Tapan Nayak for useful discussions and comments.

%\end{multicols}

\newpage
\centerline {\bf FIGURE CAPTIONS}
\vskip .1in

1) Thick solid curve represents the effective potential at
$T = 0$ with the value of the chiral field denoted by point p.
Temperature is instantaneously raised to $T_1 (< T_c)$. Dashed
curve shows the effective potential at this temperature.

2) Evolution of the temperature T (in MeV) as a function of 
proper time $\tau$ (in fm). Dotted curve shows linear heating. Solid 
and dashed curves show instantaneous 
heating with different values of $\tau_0$.

3) The evolution of $\sigma$ (in MeV). $\tau$ is in fm.

4) Different trajectories of the chiral field for the 
instantaneous heating case (solid curve in Fig.2). Figure shows 
parametric plots (upto $\tau = 10$ fm) of $\sigma$ and 
$|{\vec \pi}|$ (both in MeV).

5) Scatter plot of the projection of pionic components in
the $\pi_1 - \pi_2$ plane corresponding to Fig.4, at the stage
when T = 115 MeV. This patch covers
only about 3 \% of the area of the vacuum manifold at this stage.

\newpage

%%%%%%%%%%%%%%%%%%%%%%%%%%%%%%%%%%%%%%%%%%%%%%%%%%%%%%%%%%%%%%%%
\begin{figure}[h]
\begin{center}
\leavevmode
\epsfysize=5truecm \vbox{\epsfbox{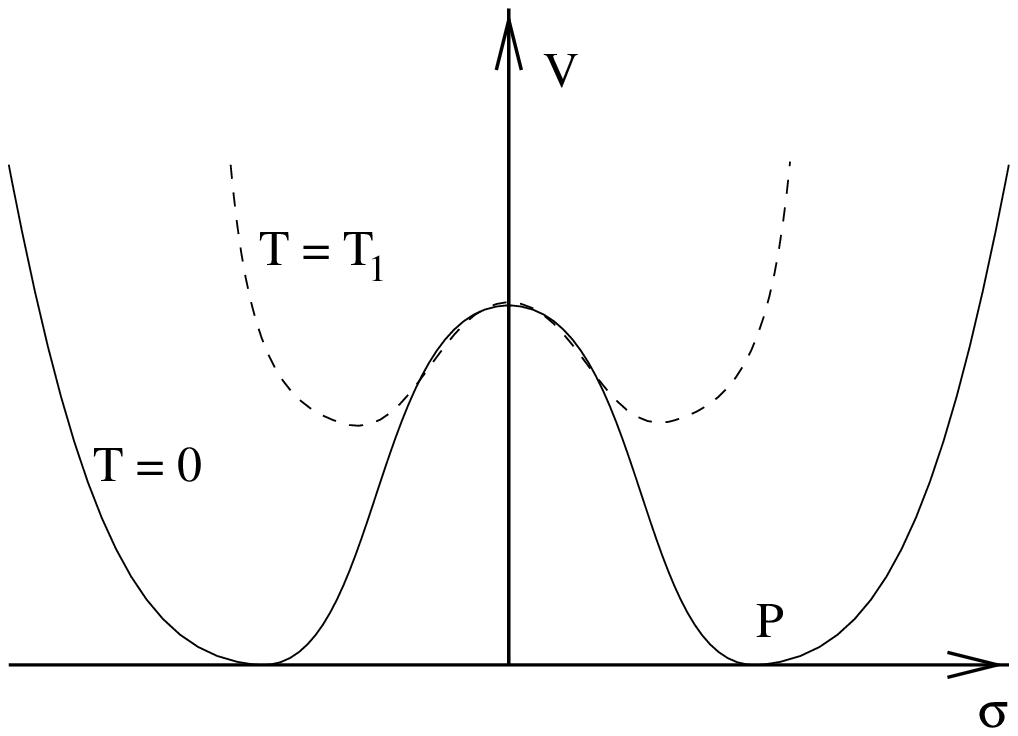}}
\end{center}
\caption{}
\label{Fig.1}
\end{figure}
%%%%%%%%%%%%%%%%%%%%%%%%%%%%%%%%%%%%%%%%%%%%%%%%%%%%%%%%%%%%%%%%%%

%%%%%%%%%%%%%%%%%%%%%%%%%%%%%%%%%%%%%%%%%%%%%%%%%%%%%%%%%%%%%%%%
\begin{figure}[h]
\begin{center}
\leavevmode
\epsfysize=20truecm \vbox{\epsfbox{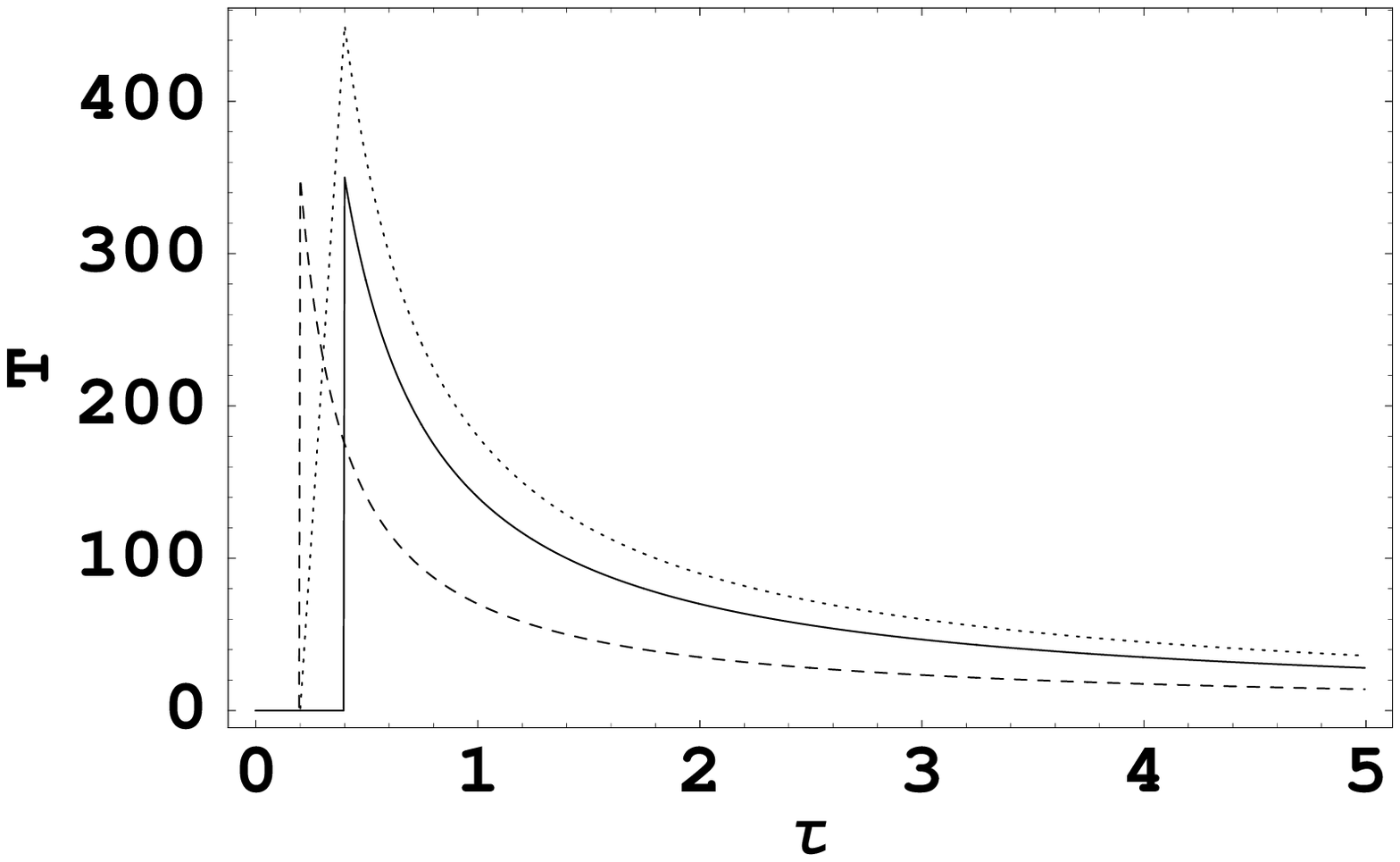}}
\end{center}
\vskip -1.5in
\caption{}
\label{Fig.2}
\end{figure}
%%%%%%%%%%%%%%%%%%%%%%%%%%%%%%%%%%%%%%%%%%%%%%%%%%%%%%%%%%%%%%%%%%
%%%%%%%%%%%%%%%%%%%%%%%%%%%%%%%%%%%%%%%%%%%%%%%%%%%%%%%%%%%%%%%%
\begin{figure}[h]
\begin{center}
\leavevmode
\epsfysize=20truecm \vbox{\epsfbox{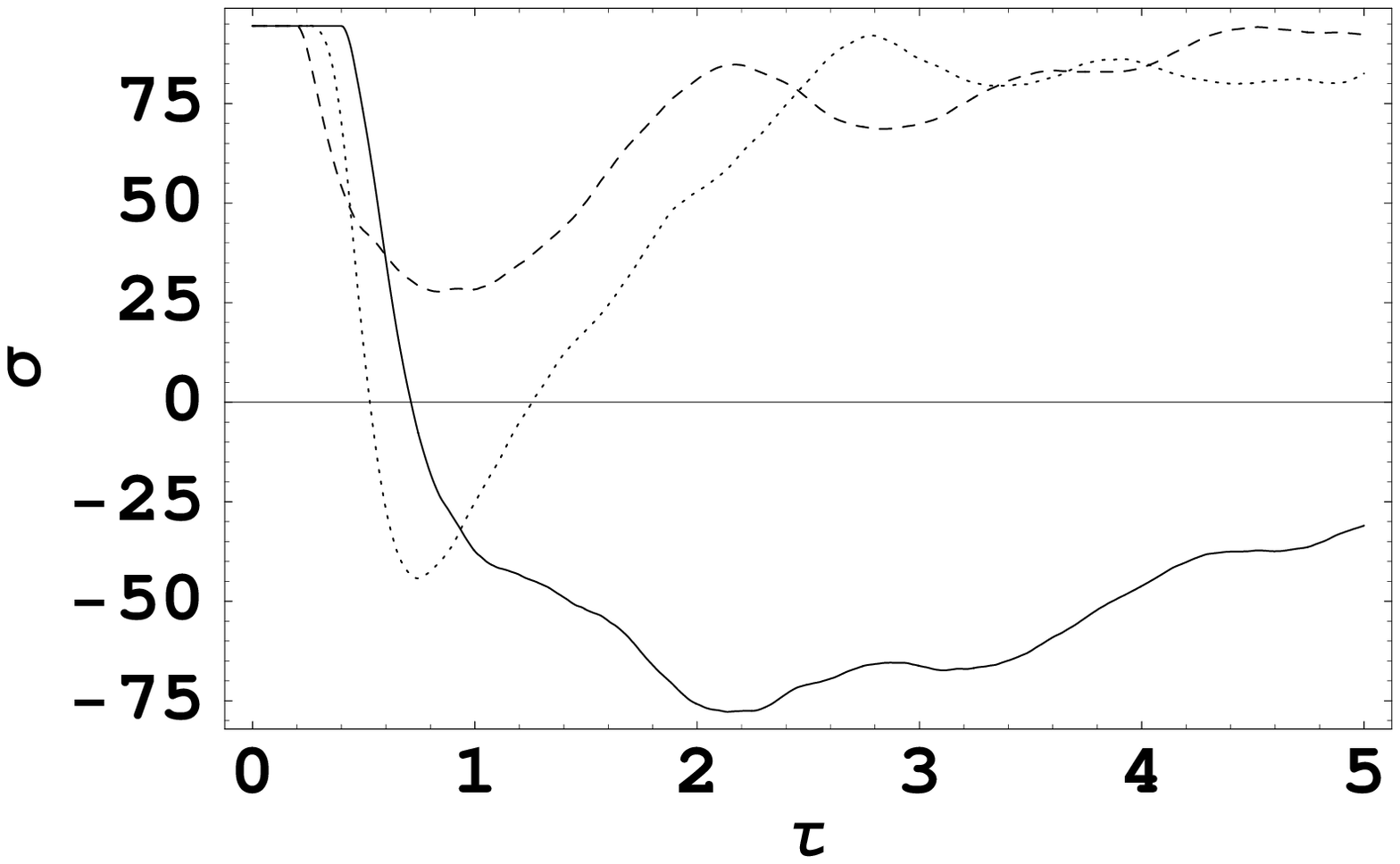}}
\end{center}
\vskip -2in
\caption{}
\label{Fig.3}
\end{figure}
%%%%%%%%%%%%%%%%%%%%%%%%%%%%%%%%%%%%%%%%%%%%%%%%%%%%%%%%%%%%%%%%%%
%%%%%%%%%%%%%%%%%%%%%%%%%%%%%%%%%%%%%%%%%%%%%%%%%%%%%%%%%%%%%%%%
\begin{figure}[h]
\begin{center}
\leavevmode
\epsfysize=20truecm \vbox{\epsfbox{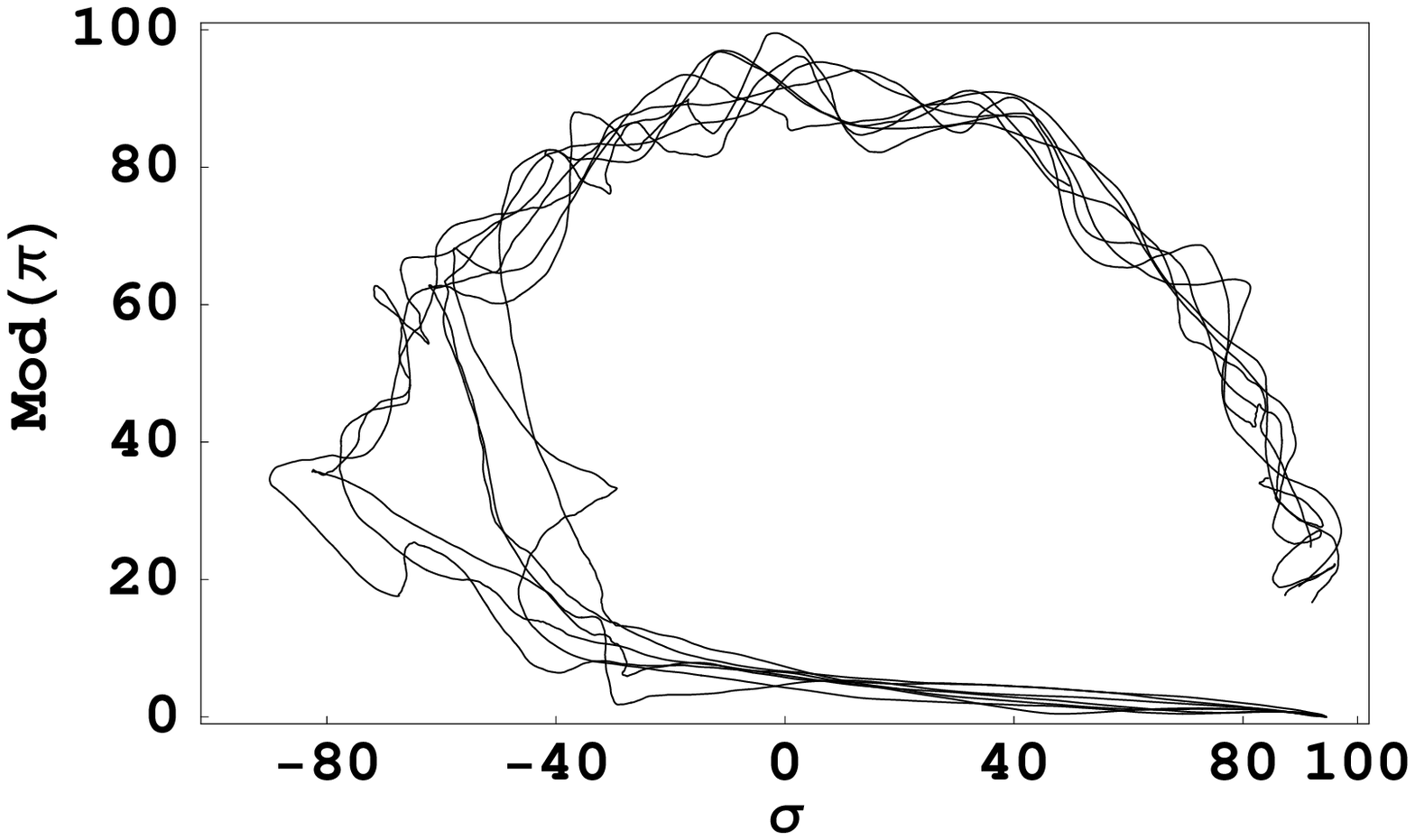}}
\end{center}
\vskip -1in
\caption{}
\label{Fig.4} 
\end{figure}
%%%%%%%%%%%%%%%%%%%%%%%%%%%%%%%%%%%%%%%%%%%%%%%%%%%%%%%%%%%%%%%%%%
%%%%%%%%%%%%%%%%%%%%%%%%%%%%%%%%%%%%%%%%%%%%%%%%%%%%%%%%%%%%%%%%
\begin{figure}[h]
\begin{center}
\leavevmode
\epsfysize=20truecm \vbox{\epsfbox{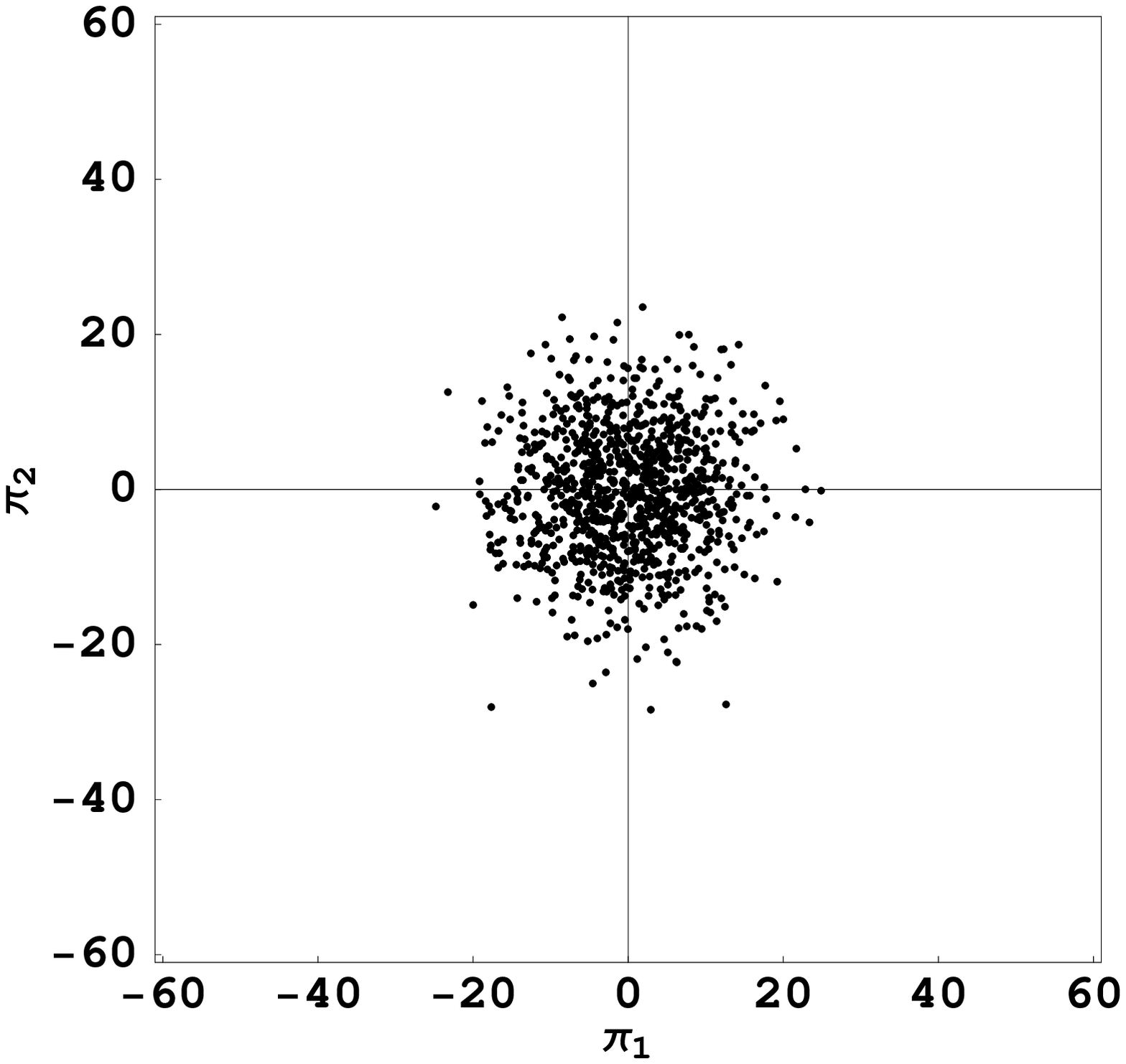}}
\end{center}
\vskip -1in
\caption{}
\label{Fig.5} 
\end{figure}
%%%%%%%%%%%%%%%%%%%%%%%%%%%%%%%%%%%%%%%%%%%%%%%%%%%%%%%%%%%%%%%%%%

\end{document}